\documentclass[10pt,preprintnumbers,twocolumn,amsmath,amssymb,nofootinbib,superscriptaddress]{revtex4-1}
\usepackage[latin1]{inputenc}
\usepackage{slashed}
\usepackage{amsmath}
\usepackage{textcomp}
\usepackage{amssymb}
\usepackage{amsfonts}
\usepackage{indentfirst}
\usepackage{color}
\usepackage{hyperref}
\usepackage[top=2cm,bottom=2cm,left=3cm,right=2cm]{geometry}
\newcommand{\nin}{\noindent}
\newcommand{\be}{\begin{equation}}
\newcommand{\ee}{\end{equation}}
\newcommand{\bea}{\begin{eqnarray}}
\newcommand{\eea}{\end{eqnarray}}

\usepackage{graphicx,graphics}
\graphicspath{{figures/}}

\begin{document}

\hfill{KCL-PH-TH/2021-01}

\title{Tunnelling and dynamical violation of the Null Energy Condition}

\author{Jean Alexandre}  
\affiliation{Theoretical Particle Physics and Cosmology, King's College London, WC2R 2LS, UK}
\author{Janos Polonyi}
\affiliation{Theory Group, CNRS-IPHC, University of Strasbourg, France}

\begin{abstract}
The Null Energy Condition is considered the most fundamental of the energy conditions, on which several key results, such as the singularity theorems, are based. The Casimir effect is one of the rare equilibrium mechanisms by which it is breached without invoking modified gravity or non-minimal couplings to exotic matter. In this work we propose an independent dynamical mechanism by which it is violated, with the only ingredients being standard (but non-perturbative) QFT and a minimally coupled scalar field in a double-well potential. 
As for the Casimir effect, we explain why the Averaged Null Energy Condition is not violated by this mechanism. 
Nevertheless, the transient behaviour could have profound impacts in Early Universe Cosmology.
\end{abstract}

\maketitle

\section{Introduction}

In General Relativity, an energy condition consists in assuming that matter satisfies "physical" properties, 
common to all forms of known matter. 
The Null Energy Condition (NEC), defined more formally below, plays an important role in Cosmology, and
for a perfect fluid in a homogeneous and isotropic Universe, it
translates to the requirement that $\rho+p\ge0$, where $\rho$ and $p$ are respectively the density and pressure of the fluid in its comoving frame. 
NEC violation $\rho+p<0$ provides a loophole in the singularity theorems that state that a collapsing universe ends at a singularity \cite{singularity}, 
and allows the possibility for a cosmological bounce \cite{bounce}. 
More generally, a fluid which violates the NEC would allow a whole new set of exotic phenomena such as traversable wormholes \cite{wormhole}.

Models violating the NEC require the introduction of non-trivial features, such as ghost condensates or Lagrangians with 
higher order derivatives, and such models have been explored extensively (see \cite{Rubakov} for a review). 
Generating the NEC violation dynamically, without introducing specific models by hand, is more difficult, although one known example is the 
Casimir effect\footnote{Out-of-equilibrium processes can also violate the NEC \cite{outofequilibrium}}, 
where vacuum fluctuations generate a negative energy density between two parallel conducting plates. The application of this phenomenon to the Early Universe is studied in \cite{ZeldoStaro}, with a confined massless scalar field inducing a cosmological expansion. 
The Casimir effect is suppressed exponentially for a massive scalar though \cite{reviewCasimir}, and can be either attractive or repulsive,
depending on the space geometry/topology.

We present here an alternative dynamical mechanism to violate the NEC,
based on non-perturbative quantum effects, arising from tunnelling between degenerate vacua 
in a finite-volume, for a massive scalar theory. The key ingredient here is the finite volume in which the field is confined, which in the Early Universe could be achieved by a shrinking causal volume. Otherwise the ingredients are prosaic - 
a minimally coupled scalar field with a Higgs-like potential.

\begin{figure}[t]
	\centering
	\includegraphics[scale=0.5]{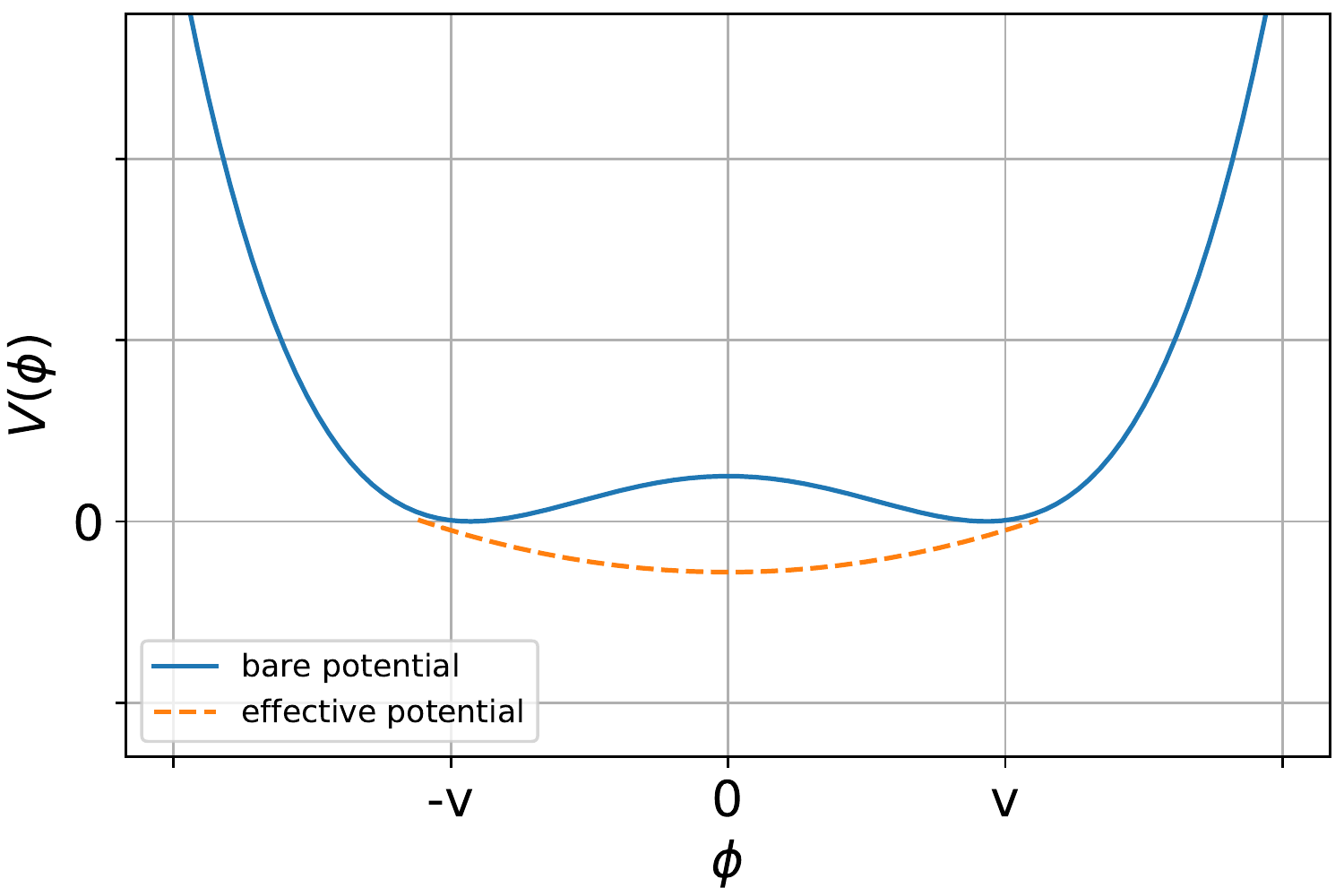}
	\caption{In finite volume, symmetry is restored by tunnelling between the two bare minima, 
	which results in a negative energy density in the ground state (taken from \cite{AC}).}
	\label{Fig:Potential}
\end{figure}

The Standard Model of Particle Physics 
assumes an ``infinite'' volume, which allows spontaneous symmetry breaking (SSB), such that the relevant partition function is a partial 
one, built on one vacuum only \cite{Tamarit}. ``Infinite'' here means large in comparison to the de Broglie wave length of particles, 
which is clearly the case in most systems. But if one allows quantum fluctuations to overlap between different vacua, 
which can happen in a finite volume, one should consider instead the full partition function, 
involving all the vacua and allowing tunnelling between these. In this case the competition between different saddle points 
leads to a convex effective potential \cite{convexity}, which restores symmetry instead of allowing SSB. This effect is illustrated in Fig. \ref{Fig:Potential}: 
the lowering of the ground state energy usually follows the enrichment of a variational state, in particular the combination of the Gaussian states, 
localised in different potential wells \cite{Kleinert}. Since energy gain from tunneling is stronger for smaller volume the NEC is expected to be violated.

The explicit calculation of the convex one-particle irreducible (1PI)
effective potential for $O(N)$-symmetric scalar theories is done in \cite{AT}, where the partition function is evaluated with a 
semi-classical approximation (ignoring fluctuations above the saddle points), 
and where the effective action $S_{eff}$ is expressed as an expansion in the classical field
(the formal large-volume limit allows the ressumation 
of all the powers of the classical field for $S_{eff}$ \cite{finiteT}).
We follow here a similar approach,
improved by considering fluctuations of the field though.

The Wilsonian approach is usually based on exact 
functional differential equations, which automatically take into account all the vacua of a theory, independently of the volume.
As a consequence the infrared effective potential is always convex, and recovers the Maxwell construction, 
featuring a flat effective potential between the two bare vacua \cite{Wilsonian}.
We also note here that the equivalence between the Wilsonian and the 1PI effective potential is valid in the limit of infinite volume only.

It was conjectured in \cite{AC} that the above finite volume effect implies a dynamical violation of the NEC, 
as a consequence of a non-trivial volume dependence (or scale factor dependence) of the action in the vicinity of the 
ground state of the dressed theory. To summarise the essence of the corresponding mechanism: 
When taking into account several saddle points, one needs to treat the quantisation four-dimensional volume $V$ 
as a parameter of the theory, which can be thought of as the volume of a box confining the scalar field.
We find then that the effective action has a non-trivial dependence on $V$, and has the form
\be
S_{eff}[\phi_0]=VU_{eff}(\phi_0,V)~,
\ee
where $U_{eff}$ is the convex effective potential evaluated at the constant classical field $\phi_0$, and also depends on $V$.
This non-extensive property was already mentioned in \cite{finiteT}, and is at the origin of a non-standard pressure, 
leading to the NEC violation. Indeed, if $\rho$ is the energy density and $p$ is the pressure for the scalar field in the 
vicinity of the true vacuum $\phi_0=0$, one has
\be
\rho+p=\frac{S_{eff}}{V}-\frac{\partial S_{eff}}{\partial V}=-V \frac{\partial U_{eff}}{\partial V}~,
\ee
which, as we will see, is negative in the regime where tunnelling occurs.

Compared to \cite{AC}, the article provides an explicit proof of the mechanism, including:\\ 
{\it(i)} an explicit calculation of the sum $\rho +p$, 
in both flat spacetime and in spatially flat Friedman-Lema\^itre-Robertson-Walker (FLRW) spacetime; \\
{\it(ii)} a quantitative justification why, in the path integral $Z$, the homogeneous saddle points dominate over the instanton;\\
{\it(iii)} an improved semi-classical approximation to estimate $Z$.

Below we start by defining the regime where tunnelling is expected to occur, and then explain in section III which 
saddle points dominate the partition function $Z$, in order to define the semi-classical approximation to calculate $Z$, and 
derive the effective action. We then generalise the calculation to a FLRW spacetime in Section IV,
where we also discuss why the Averaged NEC is not violated by the mechanism, either in static or FLRW spacetime.
Much work remains to fully elucidate this interesting effect, and we conclude by discussing future directions.

\section{Condition for tunnelling}

We start with intuitive arguments explaining in which situation one can expect tunnelling to occur, instead of SSB. 

Consider the bare potential 
\be\label{Ubare}
U_{bare}(\phi)=\frac{\lambda}{24}(\phi^2-v^2)^2~,
\ee
and the volume of quantisation $V=\int d^4x\equiv l^4$.
\begin{itemize}
 \item Quantum Mechanics point of view: The potential barrier between the vacua is $\lambda v^4/24$, corresponding to the 
 energy $\delta E=\lambda v^4 l^3/24$ in the box of volume $l^3$. The energy levels in this box are characterised by the typical gap $l^{-1}$, 
 and one can expect tunnelling to happen if this energy gap is of the order of the energy barrier, which leads to $\lambda v^4V\sim24$;
 \item Field Theory point of view: In the path integral, quadratic fluctuations of the field constant mode 
 above the bare vacua lead to a Gaussian of width $2(Vm^2/2)^{-1/2}$, where $m^2=\lambda v^2/3$. 
 Fluctuations over the two bare vacua overlap if the latter width is of the order of half the separation $2v$ of the vacua, 
 which leads to $\lambda v^4V\sim24$.
\end{itemize}
As a consequence, tunnelling between degenerate vacua should be taken into account when the parameters of the model 
satisfy the following order of magnitude 
\be\label{A}
A\equiv\lambda v^4V/24\sim1~.
\ee
If one considers the Higgs vacuum $v\simeq246$GeV and a typical perturbative coupling constant $\lambda=0.01$, the corresponding typical length
is $l\simeq10^{-17}$m, thus far larger than the Planck length, justifying a classical gravity background.

We note that the tunnelling time $\delta t$ can be approximated in two different ways, which lead to the same result for $A\sim1$:
{\it(i)} causality, which implies $\delta t\simeq l$;
{\it(ii)} uncertainty principle, which implies $\delta t\simeq1/\delta E=l/A\simeq l$.\\

\section{Effective action in the vicinity of the true vacuum}

\subsection{Saddle points}

The path integral $Z$ of this model involves both homogeneous saddle points and instantons
relating the two vacua. We show here that the latter have a negligible role in the path integral, compared to the homogeneous
saddle points, if one focuses on the true vacuum of the dressed theory. 

The definition of $Z$ requires the introduction of a source $j$, which lifts the degeneracy of the bare vacua, 
and we follow here the original arguments \cite{Coleman} describing the construction of the instanton. 
We assume a configuration $\xi$ depending on the 4-dimensional Euclidean radial coordinates $\rho=\sqrt{t^2+r^2}$, 
and described by the action
\be\label{Sinst}
S_{inst}=2\pi^2\int \rho^3 d\rho\left(\frac{1}{2}(\xi')^2+U_{bare}(\xi)+j\xi\right)~,
\ee
where the source $j$ is constant. At the zeroth-order in $j$,  
this configuration should represent a bubble of vacuum $-v$ in the environment $+v$, with a wall thickness $\delta$ (we choose $j>0$,
otherwise for energetic reasons we would consider the creation of a bubble of vacuum $+v$ in the environment $-v$).  
An approximate analytical expression for this instanton is
\be
\xi\simeq v\tanh\left(\frac{\rho-R}{\delta}\right)~~~~\mbox{with}~~~\delta=\frac{2}{v}\sqrt\frac{3}{\lambda}~,
\ee
where $R$ is the radius of the bubble, to be determined by the variational approach below. 
The action (\ref{Sinst}) has two contributions: a volume term and a surface tension. 
Keeping only the lowest order in $j$, the former is obtained from the potential energy
\be
S_{vol}\simeq Vjv+2\times 2\pi^2\frac{R^4}{4}j(-v)=jv(V-\pi^2R^4)~,
\ee
and the latter is obtained from the kinetic energy
\be
S_{surf}\simeq2\pi^2R^3\delta v^4=2\pi^2(Rv)^3\sqrt\frac{3}{\lambda}~.
\ee
The total action $S_{inst}=S_{vol}+S_{surf}$ is minimised for 
\be\label{R}
R=\frac{3v^2}{2j}\sqrt\frac{3}{\lambda}~,
\ee
which leads to 
\be\label{Sinstbis}
S_{inst}=Vjv+\frac{243\pi^2}{16\lambda^2}\frac{v^9}{j^3}~.
\ee
In the finite volume $l^4$, the radius (\ref{R}) is at most equal to $l/2$, which leads to the following lower bound for the source,
in order to create the instanton,
\be
j\ge\frac{3v^2}{l}\sqrt\frac{3}{\lambda}~.
\ee
But because tunnelling restores symmetry, the vacuum we will focus on corresponds to a vanishing classical field, 
and therefore to a vanishing source, where the action (\ref{Sinstbis}) diverges and thus doesn't contribute to the path integral.
As a consequence in what follows we take into account homogeneous saddle points only, 
which dominate the path integral in the vicinity of the true vacuum.

\subsection{Semi-classical approximation}

The partition function of the model is estimated with a semi-classical approximation, that we define here, and we show
how different saddle points are taken into account, as a consequence of tunnelling.

Unlike \cite{AT}, where fluctuations around saddle points are neglected, we consider here the following improved semi-classical 
approximation, defined by the path integral
\be\label{Z}
Z[j]\simeq\sum_k F_k\exp\left(-V[U_{bare}(\phi_k)+j\phi_k]\right)~,
\ee
where the summation runs over the homogeneous saddle points $\phi_k$, 
and the factors $F_k$ arise from the integration over quadratic fluctuations of the field constant mode:
\be
F_k=\frac{v}{\sqrt{U_{bare}''(\phi_k)}}~.
\ee
The homogeneous saddle points satisfy
\be
U_{bare}'(\phi_k)+j=0~,
\ee
and the number of solutions depends on the amplitude of the source $j$. We thus
introduce the critical source $j_c\equiv \lambda v^3/(9\sqrt3)$ to distinguish two regimes:\\

\nin\underline{$|j|\geq j_c$} In this case there is only one homogeneous saddle point 
\be
\phi_0=-\mbox{sign}(j)\frac{2v}{\sqrt3}\cosh\Big(\frac{1}{3}\cosh^{-1}(|j/j_c|)\Big)~,
\ee
and we expected from the usual 1PI construction that corrections to the bare potential should be perturbative. 
Since we work here with finite volume though, it is interesting to check that corrections are indeed small.
For a constant source the functional derivative $\delta/\delta j$ is replaced by $\partial/\partial(Vj)$,  
and the classical field is 
\be
\phi_c=-\frac{1}{ZV}\frac{\partial Z}{\partial j}=\phi_0+\frac{3\phi_0}{V(3\phi_0^2-v^2)}\frac{\partial\phi_0}{\partial j}~,
\ee
where the term proportional to $\partial\phi_0/\partial j$ arises from the fluctuation factor $F_0$. An expansion around $j_c$ gives
\be
\phi_c=-\mbox{sign}(j)\frac{2v}{\sqrt3}\left(1-\frac{1}{12A}\right)+{\cal O}(j-j_c)~,
\ee
whereas the tree-level approximation (with $F_0=1$) would give
\be
\phi_c^{tree}=-\mbox{sign}(j)\frac{2v}{\sqrt3}+{\cal O}(j-j_c)~.
\ee
Thus one can see that the correction $(12A)^{-1}$ is small compared to 1 in the regime $A\sim1$ we are interested in, 
and we neglect corrections to the bare theory for $|j/j_c|\ge1$, or equivalently $|\phi_c|\ge2v/\sqrt3$.\\

\nin\underline{$|j|\leq j_c$}, there are two homogeneous saddle points 
\bea
\phi_1&=&\frac{2v}{\sqrt3}\cos\Big(\frac{\pi}{3}-\frac{1}{3}\cos^{-1}(j/j_c)\Big)\\
\phi_2&=&\frac{2v}{\sqrt3}\cos\Big(\pi-\frac{1}{3}\cos^{-1}(j/j_c)\Big)~,\nonumber
\eea
and the partition function (\ref{Z}), normalised so that $Z(0)=1$, is expanded in the source  
\be\label{Zexp}
Z=1+\frac{12j^2}{v^6\lambda^2}(117/32-6A+24A^2)+\cdots
\ee
where dots represent higher orders in $j$. The classical field is then 
\be\label{phic}
\phi_c=-\frac{117/32-6A+24A^2}{\lambda v^2A}~j+\cdots~
\ee
and we note that there is a one-to-one mapping between the source and the classical field.
Hence this important feature of the 1PI quantisation is not modified in the presence of several saddle points, 
as long as one keeps a finite volume. Note that we keep only the lowest order in $j$ since, as explained above, 
the instanton saddle point should in principle be taken into account for large source.

\subsection{Effective action and NEC violation}

As we explain here, the effective potential is drastically different from the bare potential for $|j/j_c|\leq1$, 
or equivalently for $|\phi_c|\leq2v/\sqrt3$, as a result of the competition of two saddle points.

Based on the expansion (\ref{phic}), we express the source $j$ as a function of $\phi_c$, 
and the effective potential satisfies, for $|\phi_c|\leq2v/\sqrt3$,
\be
\frac{\partial}{\partial\phi_c}U_{eff}(\phi_c)=-j\simeq\frac{\lambda v^2 A\phi_c}{117/32-6A+24A^2}~.
\ee
We integrate the above expression  
by matching the potentials at the boundaries of the two regimes defined above 
\be
U_{eff}(\pm 2v/\sqrt3)=U_{bare}(\pm 2v/\sqrt3)=\frac{\lambda v^4}{216}~,
\ee
such that, up to higher orders in $\phi_c$,  
\be\label{Ueff}
U_{eff}(\phi_c)=\frac{\lambda v^4}{216}+\frac{\lambda v^4A[(\phi_c/v)^2-4/3]}{117/16-12A+48A^2}+\cdots
\ee
Hence the effective theory is indeed described by a convex potential, its ground state is $\phi_c=0$, and its action for a homogeneous field is 
\bea
S_{eff}(\phi_c)&=&VU_{eff}(\phi_c)\\
&=&\frac{A}{9}+\frac{24A^2[(\phi_c/v)^2-4/3]}{117/16-12A+48A^2}+\cdots\nonumber
\eea
The sum of the energy density $\rho$ and pressure $p$ in the ground state is therefore
\bea\label{rho+p}
\rho+p&=&\frac{S_{eff}(0)}{V}-\frac{\partial S_{eff}(0)}{\partial V}\\
&=&\frac{64\lambda v^4A(39-256A^2)}{9(39-64A+256A^2)^2}~,\nonumber
\eea
which is negative in the regime of interest $A\sim1$ (note that the denominator never vanishes).
The NEC is violated in the ground state because of the non-trivial volume dependence of the effective action, 
which is not extensive for $|\phi_c|\leq 2v/\sqrt3$.
We finally note that, although the regime where tunnelling occurs is $A\sim1$, the formal limit $A\to\infty$ leads to a flat 
effective potential (\ref{Ueff}), which corresponds to the so-called Maxwell construction.

\section{Spatially-flat FLRW spacetime}

\subsection{Dynamical NEC violation}

The previous results are generalised here to an expanding spacetime, and we explain how to take into account the time-dependent  
scale factor in the derivation of the effective potential. 

In a curved space time, one needs to define the concept of volume consistently, 
which can be done by considering periodic space coordinates, with cell-volume $l^3$. 
The comoving space volume is then $a^3(t)l^3$ and the 4-d volume is
\be
\int d^4x\sqrt{g}=l^3\int_{t_0}^{t_1} dt'~a^3(t')~,
\ee
where $t_1-t_0\simeq l$. We then define the time  $t$ such that
\be
la^3(t)\equiv\int_{t_0}^{t_1} dt'~a^3(t')~,
\ee
where $a(t)$ corresponds to an average of $a(t')$ for $t_0\leq t'\leq t_1$. For each comoving time $t$, 
quantisation is therefore done in the time interval $[t_0,t_1]$, where we assume that the scale factor 
remains approximately constant, allowing the use of equilibrium field theory. This approximation is even more justified 
in the vicinity of a cosmological bounce, where $\dot a=0$, which is the regime motivating \cite{AC}.

For a field minimally coupled to the metric, the Euclidean bare action for the saddle points is 
\bea
S_{bare}(\phi_k)&=&l^3\int_{t_0}^{t_1} dt'~a^3(t')U_{bare}(\phi_k)\\
&=& l^4a^3(t)U_{bare}(\phi_k) ~,\nonumber
\eea
and the tunnelling condition is 
\be
\tilde A(t)\equiv\lambda(vl)^4a^3(t)/24\simeq1~.
\ee
We then follow the same steps as those in flat space time, but the effective potential is now a function of the scale factor $a(t)$, 
through the replacement $V\to l^4a^3(t)$. The energy density in the vacuum is then
\bea
\rho(t)&=&\frac{2}{\sqrt{g}}\frac{\delta }{\delta g_{00}(t)}\int d^3xdt'\sqrt{g}~U_{eff}(t',0)\\
&=&\frac{\lambda v^4(39-1600\tilde A+256\tilde A^2)}{216(39-64\tilde A+256\tilde A^2)}~,\nonumber
\eea
and the sum $\rho+p$ can be found from the continuity equation 
\bea\label{rho+pFLRW}
&&\rho+p=-\frac{\dot\rho}{3H}=-\tilde A\frac{\partial\rho}{\partial\tilde A}\\
&=&\frac{64\lambda v^4\tilde A(39-256\tilde A^2)}{9(39-64\tilde A+256\tilde A^2)^2}~,\nonumber
\eea
where $H\equiv\dot a/a$. As expected, the expression for $\rho+p$ is similar to the one obtained in flat spacetime
(\ref{rho+p}), and the NEC is violated
as a consequence of the non-trivial dependence of the effective potential on the scale factor.

\subsection{Relevance in the Early Universe}

The effect of tunnelling could be of interest for the Early Universe, mainly in the situation where the latter starts to collapse.
In this case $\dot a<0$ and the comoving space volume decreases, until the regime where $\tilde A\sim1$ is reached, and 
tunnelling switches on. From eq.(\ref{rho+pFLRW}) we have then
\be
\rho+p\simeq-0.03\lambda v^4~,
\ee
and the scalar field ground state acts as a fluid with equation of state $w\equiv p/\rho\simeq0.1$. 
This justifies the approximation used in \cite{AC}, where $p$ is neglected compared to $\rho$.
Assuming the coexistence of this fluid and a cosmological constant with energy density $\rho_0$ and 
pressure $p_0=-\rho_0$, the Friedmann equations read
\bea
H^2&=&\frac{\kappa}{3}(\rho_0+\rho)\\
\dot H+H^2&=&-\frac{\kappa}{6}(\rho_0+\rho+3p_0+3p)~,\nonumber
\eea
where $\kappa\equiv8\pi G$. The expression for $H^2$ is consistent only if $\rho_0\ge|\rho|$, which is assumed here. 
The evolution equation for the scale factor is then
\be\label{Hdot}
\dot H=-\frac{\kappa}{2}(\rho+p)\simeq0.015\kappa\lambda v^4>0~,
\ee
such that the NEC-violating fluid induces a cosmological bounce, in the situation of an initial contraction.
The following expansion will eventually suppress tunnelling effect: 
the scalar field will then be subjected to SSB and play the role of a usual matter component.

\subsection{Averaged NEC}

We have shown that tunnelling leads to the NEC violation in two cases: static spacetime and FLRW spacetime, 
by calculating the sum $\rho +p$. A more general inequality for the NEC is 
\be
T_{\mu\nu}n^\mu n^\nu\ge0~,
\ee
at every point of spacetime and for any null vector $n^\mu$, where $T_{\mu\nu}$ is the energy momentum tensor for matter.
A weaker energy condition corresponds to the Averaged NEC (ANEC), which consists in the inequality
\be\label{ANEC}
\int d\lambda~T_{\mu\nu}n^\mu n^\nu~\ge0~,
\ee
where the integral runs along a null geodesic with tangent $n^\mu$. 
In order to discuss the ANEC, we consider here the two situations mentioned in this article: the static case and the FLRW metric. \\

$\bullet$ {\it Static case}\\ 
Here we consider a fixed box which confines the scalar field. This assumes that an external environment should be present,
in order to maintain the box structure. In this situation, it is argued in studies involving the Casimir effect, that the material from which the 
mirrors are made give a positive contribution to the integral (\ref{ANEC}), which compensates the negative energy density of the scalar field 
between the mirrors, in such a way that the ANEC is satisfied \cite{Sopova}. This is also true 
in the case where one of the mirrors has a hole \cite{Graham1}. A similar situation is obtained with an external
potential, which also confines the scalar field \cite{Graham2}: the energy associated to the potential is expected to compensate the negative energy of
the scalar field within the confined space. 
These works do not consider self-interacting fields though, whereas tunnelling on which the present article is based necessitates self-interactions. 
As a consequence, although one might expect a similar behaviour for self-interacting fields, the conclusion regarding the ANEC in our case 
requires more studies, which we leave for further works.

We note that \cite{Hartman} does consider self-interactions, and shows that the ANEC should be satisfied, 
provided the model is Lorentz-symmetric, unitary and renormalisable. 
The argument is based on the short distance properties of the dynamics, however the finite volume violates Lorentz symmetry at finite scale. 
Therefore here again, the effect described in the present article does not fit in a known context.\\

$\bullet${\it FLRW metric}\\ 
In this situation the conclusion regarding the ANEC is more straightforward. 
Here we consider instead periodic boundary conditions, for which is has been shown that the ANEC can 
be violated \cite{Fewster}. In our case though, because the scale factor will change dynamically in response to the fluid, 
the NEC-violating effect is only temporary, and the ANEC is not violated. Indeed,
we can see from eq.(\ref{Hdot}) that the effect acts to provide a bias towards expansion, 
which would eventually invalidate the finite volume condition $\tilde A\sim1$.

\section{Conclusion}

We have described how NEC violation arises naturally from tunnelling between different local minima in a finite volume, 
stressing the non-perturbative nature of this phenomenon. More specifically, the essential reason for NEC violation is 
the non-extensive feature of the effective action, as a result of several saddle points competing in the path integral.

Our work motivates the study of a number of challenging technical questions. 

Firstly, the work should be extended to the study of tunnelling in real time \cite{realtime}.
The naive extension of the 
semi-classical approximation to Minkowski spacetime leads to a complex effective action, which should lead, among other features, 
to the tunnelling rate in this model. Both the finiteness of the volume and the special role of time, played in the tunneling process, suggest that if there are dominant configurations to the path integral then they are not $O(4)$ invariant. The appropriate instanton configurations allow us to explore the double limit where the spatial size tends to infinite and the symmetry breaking external source is removed, a necessary ingredient of describing a phase transition. 

A different direction to explore is the potential relation between the Casimir effect and the present non-perturbative effect. Both 
involve finite volume and quantum fluctuations, and so it is natural to consider if a mapping between both effects can be determined. 
However, the mechanisms are clearly different, since the one presented here requires non-perturbative effects for a self-interacting field.

We also note that, in addition to Wilsonian approach, stochastic quantisation might also shed light on the mechanism described here. 
The latter approach has been used to show that curved space time has a non-trivial effect on the two-point correlation function for a 
scalar field in a double-well potential \cite{Rajantie}.

Finally, this novel mechanism for violating the NEC has the potential to open up a range of exciting research avenues. Beyond the Early Universe Cosmology application that motivated this work, one could consider whether such an effect would have an experimental signature in certain condensed matter systems. It was already suggested that phase transitions in the Early Universe could have analogues in the lab \cite{Zurek}, 
in experiments involving $^4$He. Similar experiments involving $^3$He are also discussed \cite{Brumfield}, and one 
might be able to reproduce the NEC-violation effect presented here in a physical laboratory volume.\\

\nin{\it Acknowledgements} \\
JA would like to thank K. Clough and A. Tsapalis for advice and discussions. 
The work of JA is supported by the STFC grant ST/P000258/1.

\end{document}